\documentclass[%
 notitlepage,
 twocolumn,
superscriptaddress,
 amsmath,
 amssymb,
 aps,
 pra,
]{revtex4-1}
\usepackage{amsfonts}
\usepackage{mathtools}
\usepackage{graphicx}
\usepackage{physics}
\usepackage{bm}
\usepackage{soul}
\usepackage{color}
\usepackage{xcolor}
\usepackage{dsfont}
\usepackage{hyperref}
\hypersetup{
citecolor=blue,
 colorlinks=true,
 urlcolor=magenta,
}

\def\kr{k_{\rm R}}                            			
\def\Er{E_{\rm R}}                            			
\def\Rb87{^{87}\mathrm{Rb}}                             
\def\K40{^{40}\mathrm{K}}                    		    

\newcommand{\IOPPAS}{Institute of Physics PAS, Aleja Lotnikow 32/46, 02-668 Warszawa, Poland}

\newcommand{\VILNIUS}{Institute of Theoretical Physics and Astronomy,
	Vilnius University, Saul\.etekio 3, LT-10257, Vilnius, Lithuania}

\begin{document}
	
\title{
Magnetically generated spin-orbit coupling for ultracold atoms with
slowly varying periodic driving
}
 
\author{D. Burba}
\affiliation{\VILNIUS}
\author{M. Mackoit-Sinkevi\v{c}ien\.{e}}
\affiliation{\VILNIUS}
\author{V. Novi\v{c}enko}
\affiliation{\VILNIUS}
\author{E. Witkowska}
\affiliation{\IOPPAS}
\author{G. Juzeli\=unas}
\affiliation{\VILNIUS}

\date{\today}
	
\begin{abstract}
The spin-orbit coupling (SOC) affecting the center of mass of ultracold atoms can be simulated using a properly
chosen periodic sequence of magnetic pulses. Yet such a method is generally accompanied
by micro-motion which hinders a precise control of atomic dynamics
and thus complicating practical applications.
Here we show how to by-pass the
micro-motion emerging in the magnetically induced SOC by switching
on and off properly the oscillating magnetic fields at the initial
and final times. We consider the exact dynamics of the system and demonstrate that the overall dynamics
can be immune to the micro-motion. The exact dynamics is shown to agree well with the evolution of the system described by slowly
changing effective Floquet Hamiltonian including the SOC term. The agreement is shown to be the best when the phase of the periodic driving takes
a specific value for which the effect of the spin-orbit coupling is
maximum.
\end{abstract}

\maketitle

\section{Introduction}

Spin-orbit coupling (SOC) manifests for electrons in solids \cite{Winkler03Review},
where manipulation of electron spins by SOC plays an important role
for spintronics and quantum information processing. During the last
decade there has been also a great deal of interest in SOC for ultracold atoms \cite{Ruseckas2005,Stanescu2007,Juzeliunas2008PRA,Lin2011,Goldman2014RPP,Zhai2014-review,Galitski19PT,burba2024effective}.
The SOC can lead to novel many-body phases for ultracold atoms \cite{Goldman2014RPP,Zhai2014-review,Galitski19PT}
and offer applications in areas like spintronics \cite{Vaisnav08PRL-DDT,Wilkowski22PRR}
and precision measurements \cite{Rey21PRR,Hernandez22PRL}

The SOC affecting the center of mass of ultracold atoms is usually
created by applying laser fields inducing transitions between the
atomic internal states accompanied by the recoil \cite{Ruseckas2005,Stanescu2007,Juzeliunas2008PRA,Lin2011,Goldman2014RPP,Zhai2014-review,Galitski19PT}.
This provides an effective coupling between the atomic spin and linear
momentum. Alternatively the SOC can be simulated by means of a properly
chosen periodic sequence of magnetic pulses \cite{Anderson2013,Xu2013}, and the method has been implemented for rubidium and sodium atoms \cite{Luo16SRep,Shteynas19PRL}.
Specifically, by applying to ultracold atoms an oscillating magnetic
field with a spatial gradient and an additional pulsed magnetic field,
one can simulate an effective spin-orbit coupling similar to the one
induced by laser fields. The magnetic-based approach can provide fast
and flexible changes of the system parameters, such as the recoil
momentum. This can be useful for controlling and manipulating of atomic
spin states.

The SOC created by the oscillating magnetic field is generally accompanied
by micro-motion which hinders a precise control of atomic dynamics
and thus complicates applications.
These include a fundamental study of the generation of topological states  \cite{Goldman2014RPP, Atala2014,PRXQuantum.3.030328}, subwavelength  lattices~\cite{Anderson2020,PhysRevA.107.023309}, non-trivial quantum correlations like spin sqeezing \cite{Rey21PRR, Hernandez22PRL,PhysRevB.108.104301} or indirectly Bell correlations~\cite{PhysRevLett.129.250402}. 

Here, we show how to by-pass the micro-motion emerging in the magnetically induced SOC by switching on and off properly the oscillating magnetic fields at the initial and final times. We consider the exact dynamics of the system from the initial to the final times and demonstrate that the overall dynamics can be immune to the micro-motion.  Furthermore the exact dynamics agree well with the evolution of the system described by the slowly changing effective Floquet Hamiltonian which contains the SOC term. The agreement is the best when the phase of the periodic driving takes a specific value for which the effect of the spin-orbit coupling is maximum. In that case, the first-order effective Floquet  Hamiltonian vanishes and the zero-order Floquet Hamiltonian is correct up to the second-order expansion in the inverse powers of the driving frequency. In this way, our results provide evidence that the magnetically induced SOC can be generated in a controllable way without involving the micro-motion.

The reduction of the micro-motion effect opens the path for the SOC implementation in systems where the Raman coupling is difficult to apply, for example for light atoms like lithium for which the fine structure splitting responsible for the SOC is very  small. In that case the Raman transitions inducing the SOC should be very close to the excited state resonance in order to resolve the fine structure, which might be lead to significant losses. The magnetically generated SOC does not rely on the fine structure splitting and thus provides a method for creating the SOC for a wide range of atoms including the light ones.



\section{Formulation\label{sec:Formulation}}

\subsection{Hamiltonian}

We will consider spinful atoms affected by a time-dependent inhomogeneous
magnetic field. The atomic Hamiltonian can then be separated into
a spin-independent (SI) and a spin-dependent (SD) parts: 
\begin{equation}
H=H_{\mathrm{SI}}+H_{\mathrm{SD}}\,.\label{eq:H-initial}
\end{equation}
The former SI contribution includes operators for kinetic energy for
the atomic motion in the $z$ direction and spin-independent (SI) potential
$V_{\mathrm{SI}}\left(z\right)$ which can represent any SI potential,
such as a parabolic trap or an optical lattice: 
\begin{equation}
H_{\mathrm{SI}}=\frac{p_{z}^{2}}{2m}+V_{\mathrm{SI}}\left(z\right),\label{eq:H_SI}
\end{equation}
where $z$ and $p_{z}=-i\hbar\partial_{z}$ are the atomic position and
momentum operators, $m$ being the atomic mass. 

\begin{figure}[]
\begin{centering}
\includegraphics[scale=0.6]
{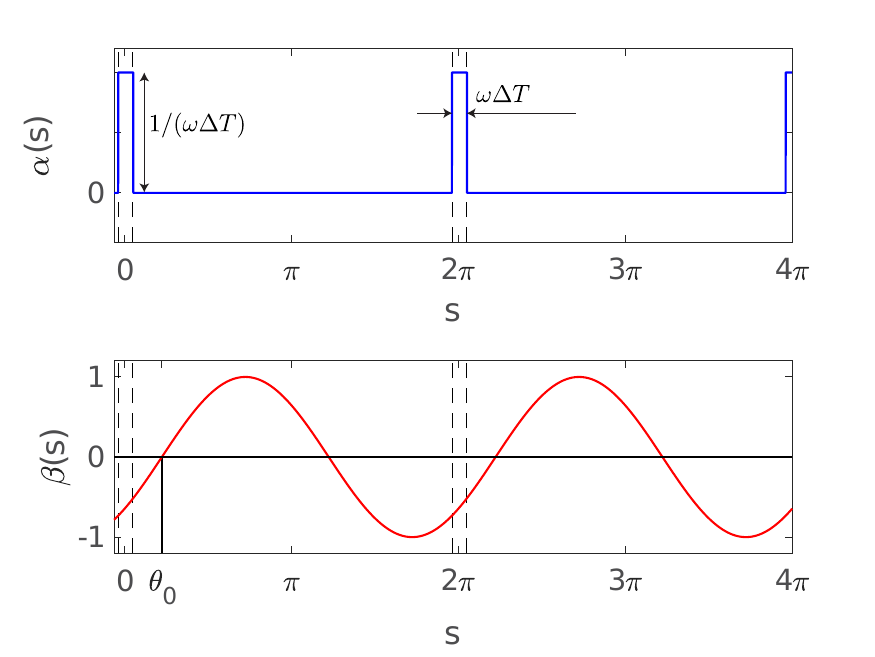} 
\par\end{centering}
\caption{\label{fig:1}The shape of the periodic functions $\alpha\left(s\right)$
and $\beta\left(s\right)$. }
\end{figure}

On the other hand, the SD terms reads

\begin{equation}
H_{\mathrm{SD}}=\omega f\left(t\right)\beta\left(\omega t\right)k_{\beta}zS_{z}+\Delta\omega_{0}S_{z}+\omega_{\alpha}g\left(t\right)\alpha\left(\omega t\right)S_{x}.\label{eq:H_SD-def}
\end{equation}
 where $S_{u}$ (with $u=x,y,z$) are the Cartesian components
of the spin operator $\mathbf{S}$. The first term in Eq.~(\ref{eq:H_SD-def}) represents the spin-dependent linear potential slope due to inhomogeneous magnetic field along the
$z$ axis. It is characterized by a slowly changing dimensionless
amplitude $f\left(t\right)$ and a periodic part $\beta\left(\omega t\right)=\beta\left(\omega t+2\pi\right)$
oscillating with a frequency $\omega=2\pi/T$. As illustrated in Fig.~\ref{fig:1},
the latter function $\beta\left(\omega t\right)$ is taken to be sinusoidal
with a tunable phase $\theta_{0}$: 
\begin{equation}
\beta\left(\omega t\right)=\sin\left(\omega t-\theta_{0}\right)\,.\label{eq:beta}
\end{equation}
The second term in Eq.~(\ref{eq:H_SD-def}) includes a possible detuning
$\Delta\omega_{0}$ between the  neighboring spin projection states. The third
term is due to a pulsed Zeeman field oriented along the $x$ axis.
The Zeeman term is characterised by a slowly changing dimensionless
amplitude $g\left(t\right)$ and a periodic part $\alpha\left(\omega t\right)=\alpha\left(\omega t+2\pi\right)$.
The latter $\alpha\left(\omega t\right)$ \textbf{ }has large non-zero
values only for a short temporal duration $\Delta T\ll T$ around
multiple integers of the driving period $t=nT$ (see Fig.~\ref{fig:1}),
where $n$ is an integer, and  each peak is normalized to unity, 
\begin{equation}
\frac{1}{2\pi}\int_{-\pi}^{\pi}\alpha\left(s\right)ds=1\,.\label{eq:alpha-normalisation}
\end{equation}
For example, $\alpha\left(\omega t\right)$ can be composed of a series
of square potentials of a temporal width $\Delta T$:

\begin{equation}
\alpha\left(\omega t\right)=\left\{
\begin{array}{cc}
\frac{1}{\omega\Delta T},\quad & -\frac{\Delta T}{2}+nT\le t<\frac{\Delta T}{2}+nT\, ,\\
0\,, & \frac{\Delta T}{2}+nT\le t<T-\frac{\Delta T}{2}+nT\,.
\end{array}\right.
\end{equation}
A specific condition how short should be the Zeeman pulses is presented
in Appendix~\ref{subsec:Estimation-of-error}. In writing Eq.~(\ref{eq:H_SD-def})
we have introduced a wave-number $k_{\beta}$ and a Rabi frequency
$\omega_{\alpha}$ characterizing, respectively, the strength of the
gradient and Zeeman fields.

The spin-dependent Hamiltonian of the form of Eq.~(\ref{eq:H_SD-def})
can be simulated using a setup which involves an oscillating quadrupole
magnetic, a strong bias magnetic field along the quantisation axis
$z$, as well as an oscillating radio frequency magnetic field
along the orthogonal ($x$) direction, see the Supplementary material
of Ref. \cite{Shteynas19PRL}. 

In the previous studies \cite{Luo16SRep,Shteynas19PRL} the gradient
and Zeeman field were considered to have constant temporal profiles, $f\left(t\right)=g\left(t\right)=1$. In that case the temporal evolution of the periodically driven quantum system is sensitive to the choice of the initial and the final times due to the micromotion \cite{Goldman2014,Eckardt2015,novi2017PRA}.
To avoid the effect of micromotion, here we introduce the slowly changing amplitudes of the oscillating gradient and Zeeman fields $f\left(t\right)$
and $g\left(t\right)$ which describe a smooth switching on and off of these fields. By setting these amplitudes to zero at the initial
and final times, we demonstrate that the overall dynamics of the periodically driven system is not sensitive to the specific choice of the initial
and final times, and is well described by the slowly changing effective Floquet Hamiltonian.

\begin{figure*}[]
\includegraphics[width=0.9\textwidth]
{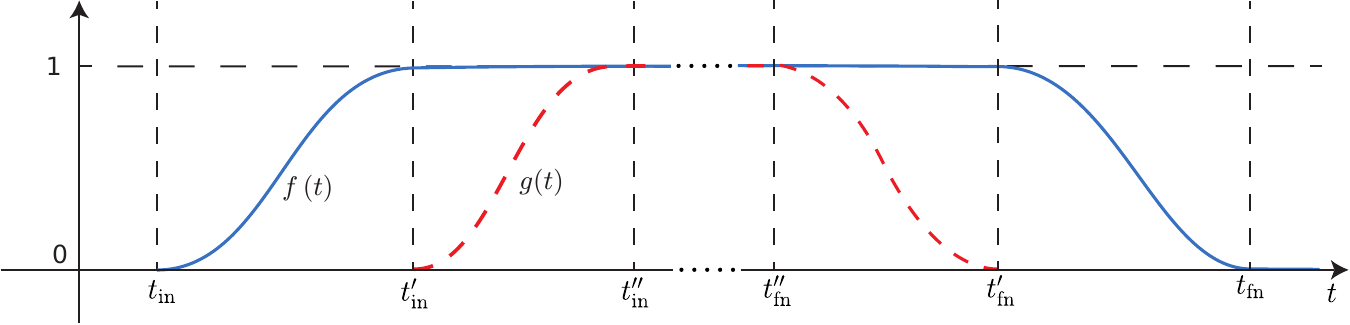} 
\caption{\label{fig:6}Schematic representation of switching on and off of
the slowly varying amplitudes $f\left(t\right)$ and $g\left(t\right)$
of the gradient and Zeeman fields represented by the blue solid and
red dashed lines, respectively.}
\end{figure*}

We will consider the following timing of the Zeeman and the gradient magnetic fields. Initially both fields are zero: $g\left(t\right)=f\left(t\right)=0$
for $t\le t_{\mathrm{in}}$. The amplitude $f\left(t\right)$ of the
gradient field is ramped up slowly from $f\left(t_{\mathrm{in}}\right)=0$
at the initial time $t_{\mathrm{in}}$ to a saturation value $f\left(t_{\mathrm{in}}^{\prime}\right)=1$
at the time $t=t_{\mathrm{in}}^{\prime}$, as illustrated schematically
in Fig.~\ref{fig:6}. During the time interval $t_{\mathrm{in}}<t<t_{\mathrm{in}}^{\prime}$
the amplitude $g(t)$ of the pulsed Zeeman field remains zero and
is ramped up during the next time interval $t_{\mathrm{in}}^{\prime}<t<t_{\mathrm{in}}^{\prime\prime}$
after the saturation of $f\left(t\right)$ is reached, as one can
see Fig.~\ref{fig:6}. The amplitudes are constant $f(t)=g(t)=1$
for $t_{\mathrm{in}}^{\prime\prime}<t<t_{\mathrm{fn}}^{\prime\prime}$
and subsequently are ramped down in the opposite order. Specifically,
the amplitude $g(t)$ is ramped down first at $t_{\mathrm{fn}}^{\prime\prime}<t<t_{\mathrm{fn}}^{\prime}$
and finally the amplitude $f(t)$ goes to zero at $t_{\mathrm{fn}}^{\prime}<t<t_{\mathrm{fn}}$,
as illustrated in Fig.~\ref{fig:6}. The implications of such a timing
for the ramping up and down of the periodic perturbation will be discussed
next.

\subsection{Elimination of the spin-dependent potential slope\label{subsec:Elimination-spin-dependent-slope}}

The multiplier $\omega$ in the first term of Eq.~(\ref{eq:H_SD-def})
reflects the fact that by increasing the driving frequency the amplitude
of inhomogeneous magnetic field is also increased. On the other hand,
we are interested in the high-frequency limit where the frequency
of the periodic driving $\omega$ exceeds all other characteristic
frequencies  featured in the Hamiltonian.  This is not the case for the spin-dependent potential
slope $\omega f\left(t\right)\beta\left(\omega t\right)k_{\beta}zS_{z}$, so this term will be eliminated in the Hamiltonian (\ref{eq:H_SD-def})
 via a time-dependent unitary transformation 
\begin{equation}
\tilde{U}_{z}\left(t\right)=\exp\left(-i\frac{z}{\hbar}k_{\beta}S_{z}\gamma\left(t\right)\right)
\,,\label{eq:U_z-tilde}
\end{equation}
where
\begin{equation}
\gamma\left(t\right)=\omega\intop_{t_{\mathrm{in}}}^{t}f\left(t_{1}\right)\beta\left(\omega t_{1}\right)dt_{1}\,.
\end{equation}
Here the lower integration limit is taken to be the initial time $t_{\mathrm{in}}$,
so that 
\begin{equation}
\gamma\left(t_{\mathrm{in}}\right)=0\quad\mathrm{and}\quad\tilde{U}_{z}\left(t_{\mathrm{in}}\right)=1\,.\label{eq:U_z-tilde-t_in}
\end{equation}
Thus the original and transformed representations coincide at the
initial time $t=t_{\mathrm{in}}$. Both representations coincide also
at the final time $t_{\mathrm{fn}}$ provided $f\left(t\right)$ is
a smooth function changing little within the driving period $T=2\pi/\omega$.
Indeed in that case the function $\gamma\left(t\right)\equiv\gamma\left(\omega t,t\right)$
can be expanded as (see Appendix~\ref{subsec:Function-gamma}) 
\begin{align}
\gamma\left(\omega t,t\right)&=-f\left(t\right)\cos\left(\omega t-\theta_{0}\right)+\frac{f^{\prime}\left(t\right)}{\omega}\sin\left(\omega t-\theta_{0}\right) \nonumber \\
&+\frac{f^{\prime\prime}\left(t\right)}{\omega^{2}}\cos\left(\omega t-\theta_{0}\right)+\ldots\,,\label{eq:gamma-rez-2}
\end{align}
so, using $f\left(t_{\mathrm{fn}}\right)=f^{\prime}\left(t_{\mathrm{fn}}\right)=f^{\prime\prime}\left(t_{\mathrm{fn}}\right)=\ldots=0$,
one finds that 
\begin{equation}
\gamma\left(t_{\mathrm{fn}}\right)=0\quad\mathrm{and}\quad\tilde{U}_{z}\left(t_{\mathrm{fn}}\right)=1\,.\label{eq:U_z-tilde-t_in-1}
\end{equation}
 As the amplitude $f\left(t\right)$ changes little within the driving period  ($f^{\prime}\left(t\right)/\omega\ll f\left(t\right)$,
$f^{\prime\prime}\left(t\right)/\omega\ll f^{\prime}\left(t\right)$,
etc.), for the present purposes it is sufficient to keep only
the zero order term in Eq. (\ref{eq:gamma-rez-2}), giving
\begin{equation}
\gamma\left(\omega t,t\right)\approx-f\left(t\right)\cos\left(\omega t-\theta_{0}\right)\,.\label{eq:gamma-rez-2-approx}
\end{equation}

The transformed Hamiltonian $\tilde{H}\left(t\right)=\tilde{U}_{z}^{\dagger}H\tilde{U}_{z}-i\hbar\tilde{U}_{z}^{\dagger}\partial_{t}\tilde{U}_{z}$
reads
\begin{align}
\tilde{H}\left(\omega t,t\right)
&=H_{\mathrm{SI}}-\frac{p_{z}k_{\beta}}{m}S_{z}\gamma\left(\omega t,t\right) \nonumber\\
&+\omega_{\alpha}g\left(t\right)\alpha\left(\omega t\right)\tilde{S}_{x}\left(z,\omega t,t\right)+\frac{k_{\beta}^{2}}{2m}S_z^2\gamma^2\left(\omega t,t\right)
\,,\label{eq:H-tilde-1}
\end{align}
where the transformed spin operator $\tilde{S}_{x}\left(z,t\right)=\tilde{U}_{z}^{\dagger}S_{x}\tilde{U}_{z}$
describes spin rotation around the $z$ axis: 
\begin{equation}
\tilde{S}_{x}\left(z,\omega t,t\right)=\cos\left(zk_{\beta}\gamma\left(\omega t,t\right)\right)S_{x}-\sin\left(zk_{\beta}\gamma\left(\omega t,t\right)\right)S_{y}\,.\label{eq:S_x-tilde}
\end{equation}
The periodic function $\alpha\left(\omega t\right)$ multiplying $\tilde{S}_{x}\left(z,t\right)$
in the Hamiltonian of Eq.~(\ref{eq:H-tilde-1}) is non-zero only
in a narrow vicinity of multiple integers of the driving period $T=2\pi/\omega$.
Therefore one can replace $\gamma\left(t\right)\equiv\gamma\left(\omega t,t\right)$
by $\gamma\left(0,t\right)=-f\left(t\right)\cos\theta_{0}$ in Eq.
(\ref{eq:S_x-tilde}) for $\tilde{S}_{x}\left(z,t\right)$, see Appendix~\ref{subsec:Estimation-of-error}
for estimating an error. Furthermore, since the function $f\left(t\right)$
reaches its saturation value $f\left(t\right)=1$ when $g\left(t\right)$
is still zero, one can put $f\left(t\right)=1$ in $\gamma\left(0,t\right)$
entering $\tilde{S}_{x}\left(z,t\right)$, so that one can make the
following replacement in the Hamiltonian of Eq.~(\ref{eq:H-tilde-1}) 
\begin{equation}
\tilde{S}_{x}\left(z,\omega t,t\right)\rightarrow\tilde{S}_{x}\left(z,t\right)
\end{equation}
with
\begin{equation}
	\tilde{S}_{x}\left(z,t\right)=\cos\left(zk_{\beta}\cos\theta_{0}\right)S_{x}+\sin\left(zk_{\beta}\cos\theta_{0}\right)S_{y}\,.\label{eq:S_x-tilde_simplified}
\end{equation}
 The amount of spin rotation is thus determined by the the wave-number $k_{\beta}\cos\theta_{0}$ times the distance $z$.

\section{Exact and effective evolution}

\subsection{Effective Floquet Hamiltonian}

In the original Hamiltonian given by Eqs.~(\ref{eq:H-initial})-(\ref{eq:H_SD-def})
the periodic perturbation represents the spin-dependent potential
slope $\omega f\left(t\right)\beta\left(\omega t\right)k_{\beta}zS_{z}$
proportional to $\omega$. In the transformed Hamiltonian $\tilde{H}\left(\omega t,t\right)$
given by Eq.~(\ref{eq:H-tilde-1}) this term is eliminated, and the
oscillating perturbation is no longer proportional to the driving
frequency $\omega$. The atomic dynamics can then be well described
in terms of a slowly changing effective Floquet Hamiltonian $H_{\mathrm{eff}}\left(t\right)$
which can be expanded in the powers of the inverse driving frequency
$1/\omega$, a procedure known as the high frequency expansion \cite{novi2017PRA}:
\begin{equation}
H_{\textrm{eff}}\left(t\right)=H_{\textrm{eff}\left(0\right)}\left(t\right)+H_{\textrm{eff}\left(1\right)}\left(t\right)+\ldots,\label{eq:H_eff_expansion}
\end{equation}
where the $n$th term $H_{\textrm{eff}\left(n\right)}$ is proportional
to $\omega^{-n}$. We will restrict to the first two terms given by
\begin{align}
H_{\textrm{eff}\left(0\right)} & =H^{\left(0\right)}\left(t\right),\label{eq:H_eff_0}\\
H_{\textrm{eff}\left(1\right)} & \left(t\right)=\frac{1}{\hbar\omega}\sum_{l=1}^{\infty}\frac{1}{l}\left[H^{\left(l\right)}\left(t\right),H^{\left(-l\right)}\left(t\right)\right],\label{eq:H_eff_1}
\end{align}
where $H^{\left(l\right)}\left(t\right)$ are slowly changing operators
featured in the Fourier expansion of the time-periodic Hamiltonian
$H\left(\omega t,t\right)=H\left(\omega t+2\pi,t\right)$ with respect
to the first argument $\omega t$: 
\begin{equation}
H\left(\omega t,t\right)=\sum_{l=-\infty}^{\infty}H^{\left(l\right)}\left(t\right)e^{il\omega t}\,.\label{eq:H-periodic-expansion}
\end{equation}
Since $\gamma\left(\omega t,t\right)$ given by Eq.~(\ref{eq:gamma-rez-2})
is expanded in the inverse powers of $\omega$, the Fourier components
$H^{\left(l\right)}\left(t\right)$ can  also be expanded in the
powers of $1/\omega$, i.e. $H^{\left(l\right)}\left(t\right)=H_{0}^{\left(l\right)}\left(t\right)+H_{1}^{\left(l\right)}\left(t\right)+\cdots$.
In the present situation $H_{1}^{\left(0\right)}\left(t\right)=0$,
so it is sufficient to keep only the leading term of $\gamma\left(\omega t,t\right)$  given by Eq.\eqref{eq:gamma-rez-2-approx}
when considering $H_{\textrm{eff}\left(0\right)}\left(t\right)$ and
$H_{\textrm{eff}\left(1\right)}\left(t\right)$.

The Fourier component $H^{\left(0\right)}\left(t\right)$ providing
the zero-order effective Floquet Hamiltonian $\tilde{H}_{\mathrm{eff}\left(0\right)}\left(t\right)$
is obtained by averaging the Hamiltonian $\tilde{H}\left(\omega t,t\right)$
with respect to the rapidly changing argument $\omega t$. According
to Eq.~(\ref{eq:gamma-rez-2}), the function $\gamma\left(\omega t,t\right)$
averages to zero $\left(2\pi\right)^{-1}\int_{0}^{2\pi}\gamma\left(s,t\right)ds=0$,
and the average of its square is $\left(2\pi\right)^{-1}\int_{0}^{2\pi}\gamma^{2}\left(s,t\right)ds=f^{2}\left(t\right)/2.$
Furthermore, according to Eq.~(\ref{eq:alpha-normalisation}), $\alpha\left(\omega t\right)$
averages to the unity over the period. Thus, using Eqs.~(\ref{eq:H-tilde-1})
and (\ref{eq:S_x-tilde_simplified}) for $\tilde{H}\left(\omega t,t\right)$,
the slowly changing zero-order effective Floquet Hamiltonian reads:\textbf{ }

\begin{equation}
\tilde{H}_{\mathrm{eff}\left(0\right)}\left(t\right)=H_{\mathrm{SI}}+\frac{g\left(t\right)\omega_{\alpha}}{2\pi}\tilde{S}_{x}\left(z,t\right)+
\frac{k_{\beta}^{2}f^{2}\left(t\right)}{4m}S_z^2
\,,\label{eq:H-tilde-eff_0}
\end{equation}
  In what follows we will consider the case of the spin 1/2 for which the Cartesian components of the spin operator read $S_{u}=\hbar\sigma_{u}/2$ (with $u=x,y,z$), where $\sigma_{u}$ are the Pauli matrices. In that case $S_z^2=\hbar^2/4$, so the last term of Eq.\eqref{eq:H-tilde-eff_0} is spin independent and represents the slowly changing shift in the origin
of energy. 
As demonstrated in Appendix~\ref{Appendix_B}, for the spin-1/2 one can make simplifications also to the first order effective Hamiltonian leading to the following result:  
\begin{equation}
H_{{\rm eff}\left(1\right)}\left(t\right)=\frac{\omega_{\alpha}\hbar k_{\beta}f\left(t\right)g\left(t\right)}{4\pi m\hbar\omega}\sin\left(\theta_{0}\right)\left( p_{z}\tilde{S}_{y} + \tilde{S}_{y} p_{z} \right)\,.
\label{eq:H_eff1}
\end{equation}
 where $\tilde{S}_{y}\equiv \tilde{S}_{y}\left(z,t\right)$ is given by
\begin{equation}
\tilde{S}_{y}\left(z,t\right)=
\cos\left(zk_{\beta}\cos\theta_{0}\right)S_{y}-
\sin\left(zk_{\beta}\cos\theta_{0}\right)S_{x}
\,.
\end{equation}
Note that the first order contribution $\tilde{H}_{\textrm{eff}\left(1\right)}\left(t\right)$
to the effective Hamiltonian reduces to zero for the most interesting
situation where $\theta_{0}=0$, in which the momentum of spin-orbit coupling $k_{\beta}\cos\theta_{0}$ is
maximum and the condition~(\ref{eq:gamma-Taylor-1}) holds
the best, as discussed below. In that case the zero-order effective
Hamiltonian $\tilde{H}_{\mathrm{eff}\left(0\right)}\left(t\right)$
describes effectively the evolution of the system up to the terms
quadratic in the inverse frequency $1/\omega$.

The operators $\tilde{H}_{\mathrm{eff}\left(0\right)}\left(t\right)$
and $\tilde{H}_{\mathrm{eff}\left(1\right)}\left(t\right)$ change
slowly in time due to slow changes of the amplitudes of the gradient
and Zeeman fields $f\left(t\right)$ and $g\left(t\right)$. The spin
rotation term in Eq.~(\ref{eq:H-tilde-eff_0}) for the effective
Hamiltonian $\tilde{H}_{\mathrm{eff}\left(0\right)}\left(t\right)$
represents the spin orbit coupling (SOC) characterised by the slowly
changing strength $g\left(t\right)\omega_{\alpha}$ and the wave-number
of the momentum transfer $k_{0}=k_{\beta}\cos\theta_{0}$. The wave-number
$k_{0}$ is determined by the phase $\theta_{0}$ between the gradient
and the pulsed Zeeman fields, like in the stationary case where $f\left(t\right)=g\left(t\right)=1$
\cite{Shteynas19PRL}. The momentum transfer is maximum and equals
to $k_{\beta}$ for $\theta_{0}=0$ when the spikes of the infrared Zeeman
field are situated at zeros of the gradient field. In that case the
condition (\ref{eq:gamma-Taylor-1}) holds best, and also there is
no first order contribution to the effective Hamiltonian, $\tilde{H}_{\mathrm{eff}\left(1\right)}\left(t\right)=0$.
On the other hand, the momentum transfer is zero for $\theta_{0}=\pm\pi/2$
when the spikes of the Zeeman field coincide with maxima/minima
of the gradient field.

\subsection{Dynamics of the system}

The overall dynamics of the state vector of the system from the initial to the final times governed by the slowly changing periodic Hamiltonian
$\tilde{H}\left(\omega t,t\right)=\tilde{H}\left(\omega t+2\pi,t\right)$
is described by the evolution operator 
\begin{equation}
U\left(t_{\mathrm{fn}},t_{\mathrm{in}}\right)={\cal T}\exp\left[-\frac{i}{\hbar}\int_{t_{\mathrm{in}}}^{t_{\mathrm{fn}}}\tilde{H}\left(\omega t^{\prime},t^{\prime}\right)\mathrm{\mathrm{d}t^{\prime}}\right]\,,\label{eq:U-general}
\end{equation}
where ${\cal T}$ signifies the time-ordering. The operator $U\left(t_{\mathrm{fn}},t_{\mathrm{in}}\right)$
can be represented in terms of the effective evolution operator $U_{\textrm{eff}}\left(t_{\mathrm{fn}},t_{\mathrm{in}}\right)$
due to the slowly changing effective Hamiltonian $\tilde{H}_{\mathrm{eff}}\left(t\right)$
and the micromotion operators $U_{\textrm{Micro}}\left(\omega t,t\right)$
and $U_{\textrm{Micro}}^{\dagger}\left(\omega t,t\right)$ calculated
at the initial and final times $t=t_{\mathrm{in}}$ and $t=t_{\mathrm{fn}}$
\cite{novi2017PRA}: 
\begin{align}
&U\left(t_{\mathrm{fn}},t_{\mathrm{in}}\right)=\nonumber \\
&U_{\textrm{Micro}}\left(\omega t_{\mathrm{fn}},t_{\mathrm{fn}}\right)U_{\textrm{eff}}\left(t_{\mathrm{fn}},t_{\mathrm{in}}\right)U_{\textrm{Micro}}^{\dagger}\left(\omega t_{\mathrm{in}},t_{\mathrm{in}}\right)\,.\label{eq:U}
\end{align}
 Here the effective evolution is given by
\begin{equation}
U_{\textrm{eff}}\left(t_{\mathrm{fn}},t_{\mathrm{in}}\right)={\cal T}\exp\left[-\frac{i}{\hbar}\int_{t_{\mathrm{in}}}^{t_{\mathrm{fn}}}\tilde{H}_{\textrm{eff}}\left(t^{\prime}\right)\mathrm{\mathrm{d}t^{\prime}}\right]\,,\label{U_eff}
\end{equation}
and the micromotion operator reads up to the terms linear in $1/\omega$:
\begin{equation}
U_{\textrm{Micro}}\left(\omega t,t\right)\approx1-\frac{1}{\hbar\omega}\sum_{m\ne0}\frac{1}{m}H^{\left(m\right)}\left(t\right)e^{im\omega t}.\label{eq:U_Micro}
\end{equation}
The operator $U_{\textrm{Micro}}\left(\omega t,t\right)$ describes
effects due to the fast changes of the Hamiltonian $\tilde{H}\left(\omega t,t\right)$.
It goes to the unity when periodic driving switches off \cite{novi2017PRA}.
Thus in the present situation the micromotion operator $U_{\textrm{Micro}}\left(\omega t,t\right)$
reduces to the unity for $t=t_{\mathrm{in}}$ and $t=t_{\mathrm{fn}}$.
In this way, the overall dynamics described by the slowly changing
effective Floquet Hamiltonian $\tilde{H}_{\mathrm{eff}}\left(t\right)=\tilde{H}_{\mathrm{eff}\left(0\right)}\left(t\right)+\tilde{H}_{\mathrm{eff}\left(1\right)}\left(t\right)+\ldots$
should reproduce well the exact dynamics governed by the exact Hamiltonian
$\tilde{H}\left(\omega t,t\right)$. An additional temporal dependence
is due to the time-dependent unitary operator $\tilde{U}_{z}\left(t\right)$
transforming the original state vector to the new representation.
Yet, the transformation $\tilde{U}_{z}\left(t\right)$ given by Eq.~(\ref{eq:U_z-tilde})
reduces to the unity at the initial and final times and thus does
not affect the overall evolution of the system from the initial to
the final times. Therefore one can consider the time evolution from
the initial to the final times governed by the transformed Hamiltonian
$\tilde{H}\left(\omega t,t\right)$, which is turn can be described
by the slowly changing effective Floquet Hamiltonian $\tilde{H}_{\mathrm{eff}}\left(t\right)$,
i.e. 
\begin{equation}
U\left(t_{\mathrm{fn}},t_{\mathrm{in}}\right)=U_{\textrm{eff}}\left(t_{\mathrm{fn}},t_{\mathrm{in}}\right)\,.\label{eq:U-1}
\end{equation}
In the next Subsection we will make sure that this is the case.

In the stationary regime where $f\left(t\right)=g\left(t\right)=1$
the effective Floquet Hamiltonian \eqref{eq:H-tilde-eff_0} becomes time-independent and is
given by for the  case of spin 1/2
\begin{equation}
\tilde{H}_{\mathrm{eff}\left(0\right)}=H_{\mathrm{SI}}+\frac{\omega_{\alpha}}{2\pi}\left[\cos\left(zk_{0}\right)S_{x}-\sin\left(zk_{0}\right)S_{y}\right]+\frac{\hbar^2 k_{\beta}^{2}}{16m}\,,\label{eq:H-tilde-eff_0-stationary}
\end{equation}
with $k_{0}=k_{\beta}\cos\theta_{0}$.

The effective Hamiltonian given by Eq.~(\ref{eq:H-tilde-eff_0})
or (\ref{eq:H-tilde-eff_0-stationary}) is analogous to the light
induced coupling between the (quasi-) spin up and down states accompanied
by the recoil $k_{0}$, like the one used to study the spin squeezing
in optical lattices \cite{Hernandez22PRL}. The Hamiltonian reduces
to the SOC involving coupling between the linear momentum $p_{x}$
and the spin component $S_{x}$ via the unitary transformation $\exp\left[i\frac{z}{\hbar}k_{0}S_{z}\right]$
\cite{Lin2011}. Note also that the effective Hamiltonian (\ref{eq:H-tilde-eff_0})
or (\ref{eq:H-tilde-eff_0-stationary}) has been derived under the
high frequency assumption implying that the driving frequency is larger
than all the frequencies associated with the time-periodic Hamiltonian
$\tilde{H}\left(\omega t,t\right)$ changing slowly within the driving
period. In that case the effective Hamiltonian reproduces very well
the exact evolution, as we will see next.

\subsection{Exact vs numerical results}\label{Exact vs numerical results}

We will compare the time evolution of the time-dependent Schroedinger equation (TDSE), calculated numerically for both cases: the exact time-periodic Hamiltonian $\tilde{H}\left(\omega t,t\right)$ and the effective Hamiltonian $\tilde{H}_{\mathrm{eff}\left(0\right)}\left(t\right)$. The two component (spinor) wavefunctions
$|\psi\left(t\right)\rangle$ and $|\phi\left(t\right)\rangle$ governed by $\tilde{H}\left(\omega t,t\right)$ and $\tilde{H}_{\mathrm{eff}\left(0\right)}\left(t\right)$,
respectively, are chosen to be  the same at the initial time, $|\phi\left(t_{\mathrm{in}}\right)\rangle=|\psi\left(t_{\mathrm{in}}\right)\rangle$.
Both spinor wave functions should be almost identical at the final
time, $|\phi\left(t_{\mathrm{fin}}\right)\rangle=|\psi\left(t_{\mathrm{fin}}\right)\rangle$,
if the high frequency conditions are met: 
$(\mathrm{i})$ $p_{z}^{2}/\left(2m\right)\ll\hbar\omega$,
$(\mathrm{ii})$ $p_{z}k_{0}/m\ll\omega$,
$(\mathrm{iii})$ $\omega_{\alpha}\ll\omega$,
$(\mathrm{iv})$ $f^{\prime}\left(t\right)\ll\omega,\quad\mathrm{and}\quad g^{\prime}\left(t\right)\ll\omega.$
Additionally, the duration of the Zeeman pulses $\Delta T$ should
be small enough compared to the driving period $T=2\pi/\omega$, so there is an extra condition following from (\ref{eq:gamma-Taylor-1})
in Appendix~\ref{subsec:Estimation-of-error}: $(\mathrm{v})$ $Lk_{\beta}\left|\omega\Delta T\sin\theta_{0}-\left[\left(\omega\Delta T\right)^{2}/2\right]\cos\theta_{0}\right|\ll1$,
where the sample length $L$ is taken to be much larger than the inverse
momentum $1/k_{\beta}$, i.e. $Lk_{\beta}\gg1$. The condition (v) is satisfied in the experiment \cite{Shteynas19PRL} where the sample length is of the order of $100\,\mu$m, the wave-number $k_{\beta}$ of the order of $\left(\mu \mathrm{m}\right)^{-1}$ and $\omega\Delta T = 0.01 \ll 1$. Note that the condition $(\mathrm{v})$
holds best if the phase difference is zero: $\theta_{0}=0$, i.e.
when the spikes of the Zeeman field $\alpha\left(\omega t\right)$
are situated at zero points of the profile $\beta\left(\omega t\right)$. In that case the condition (v) reduces to $Lk_{\beta}\left(\omega\Delta T\right)^{2}/2\ll1$.

In the numerical calculations, we will assume that the atoms are confined
in a square well with infinitely high potential boundaries at $z=\pm L/2$
and zero potential for $z\in\left[-L/2,L/2\right]$. The ramping functions
$f\left(t\right)$ and $g\left(t\right)$ are taken to have the following
form

\begin{equation}
f\left(t\right)=\frac{1}{2}\left[\mathrm{tanh}\left\{ c\left(t-\tau/2\right)\right\} +\tanh\left\{ c\left(7\tau/2+\tau^{\prime\prime}-t\right)\right\} \right]\,,
\end{equation}

\begin{equation}
g\left(t\right)=\frac{1}{2}\left[\mathrm{tanh}\left\{ c\left(t-3\tau/2\right)\right\} +\tanh\left\{ c\left(5\tau/2+\tau^{\prime\prime}-t\right)\right\} \right]\,.
\end{equation}
where $\tau^{\prime\prime}$ is the time interval between the ramping on
and off, $1/c$ is the ramping time of the periodic driving and $\tau$
is the time delay between the ramping of the functions $f\left(t\right)$
and $g\left(t\right)$. By taking $c\tau>4$ we choose $t_{\mathrm{in}}=0$,
$t_{\mathrm{in}}^{\prime}=\tau$, $t_{\mathrm{in}}^{\prime\prime}=2\tau$
and similarly $t_{\mathrm{fn}}^{\prime\prime}=\tau^{\prime\prime}+2\tau$,
$t_{\mathrm{fn}}^{\prime}=\tau^{\prime\prime}+3\tau$, $t_{\mathrm{fn}}=\tau^{\prime\prime}+4\tau$.
In that case we have $f\left(0\right)\approx g\left(\tau\right)\approx0$
and $f\left(\tau\right)\approx g\left(2\tau\right)\approx1$, as well
as $f\left(\tau^{\prime\prime}+4\tau\right)\approx g\left(\tau^{\prime\prime}+3\tau\right)\approx0$
and $f\left(\tau^{\prime\prime}+3\tau\right)\approx g\left(\tau^{\prime\prime}+2\tau\right)\approx1$,
as illustrated in Fig.~\ref{fig:6}. 

  Note that according to the condition (iv) presented at the beginning of Sec.~\ref{Exact vs numerical results}, the ramping rate $f^\prime(t)\sim c$ should be much smaller the driving frequency $\omega$. On the other hand, the ramping time should be smaller than the decoherence time $\tau_{\mathrm{decoh}}$. The latter condition can not be met in the experiment of ref.~\cite{Shteynas19PRL} in which the decoherence time is of the order of 1 ms whereas driving period is only around 10 times smaller. To satisfy the slow ramping condition, one should increase the decoherence time, which is expected to be done in the future experiments. 
In the subsequent plots displayed in Fig.~\ref{fig:1-1} and Fig.~\ref{fig:1-1-1} the ramping rate is taken to be 100 times smaller than the driving frequency, which can be applied to future experiment with the relative decoherence times $\omega \tau_{\mathrm{decoh}}$ larger than that in Ref.~\cite{Shteynas19PRL}.

\subsubsection{Comparison of exact and effective dynamics}

To compare the dynamics, we will look at the inner product between
the state vectors $|\phi\left(t\right)\rangle$ and $|\psi\left(t\right)\rangle$
evolving, respectively, by the exact time periodic Hamiltonian $\tilde{H}\left(\omega t,t\right)$
and the effective slowly changing Hamiltonian $\tilde{H}_{\mathrm{eff}\left(0\right)}\left(t\right)$:
\begin{align}
\langle\psi\left(t\right)|\phi\left(t\right)\rangle&=\int_{0}^{L}\mathrm{d}z\,\langle\psi\left(t,z\right)|\phi\left(t,z\right)\rangle \nonumber \\
&=\int_{0}^{L}\mathrm{d}z\,\sum_{s=\{\uparrow,\downarrow\}}\langle\psi\left(t,z\right)|s\rangle\langle s|\phi\left(t,z\right)\rangle\,.
\end{align}
If  the inner product $\langle\psi\left(t\right)|\phi\left(t\right)\rangle$is unity for $t=t_{\mathrm{fn}}$,
the overall dynamics of the two state-vectors is equivalent. Otherwise,
 this is not the case. Thus for numerics, one may look at $\left|\langle\psi\left(t\right)|\phi\left(t\right)\rangle\right|$
and $\textrm{arg}\left( \langle\psi\left(t\right)|\phi\left(t\right)\rangle \right) $.
Deviations of these quantities from $1$ and $0$, respectively,
signify differences between the state-vectors and thus, the non-equivalence
of the dynamics.
 We will explore this differences for state-vectors characterized by
three orthogonal initial spin polarizations.  For this we introduce the following functions:

\begin{equation}
G\left(t\right)\coloneqq\frac{1}{3}\sum_{i=\{x,y,z\}}\left|\langle\psi^{(i)}\left(t\right)|\phi^{(i)}\left(t\right)\rangle\right|\,,
\end{equation}

\begin{equation}
A\left(t\right)\coloneqq\frac{1}{3}\sum_{i=\{x,y,z\}}\mathrm{arg}\,\langle\psi^{(i)}\left(t\right)|\phi^{(i)}\left(t\right)\rangle\,,
\end{equation}
where once again, deviations  of these functions from $1$ and $0$ signify differences
between the state-vectors.

Specifically,  the spatial part of the initial state vectors is taken to be an eigenstate 
of the box potential $\Phi_{n}\left(z\right)$,  and the spin is pointing
along the $x$, $y$ and $z$ axis: 

\begin{equation}
|\psi^{(i)}_n\left(t=0\right)\rangle=\Phi_{n}\left(z\right)|i\rangle\,,\quad i\in\left\{ x,y,z\right\} \,,
\end{equation}
where:

\begin{equation}
\Phi_{n}\left(z\right)=\sqrt{\frac{2}{L}}\sin\left(\frac{\pi n}{L}z\right)\,,
\end{equation}
and
\begin{equation}
|x\rangle=\frac{1}{\sqrt{2}}\begin{pmatrix}1\\
1
\end{pmatrix}\,,\ |y\rangle=\frac{1}{\sqrt{2}}\begin{pmatrix}1\\
\mathrm{i}
\end{pmatrix}\,,\ |z\rangle=\begin{pmatrix}1\\
0
\end{pmatrix}\,,
\end{equation}

\begin{figure}[]
\begin{centering}
\includegraphics[width=1\columnwidth]
{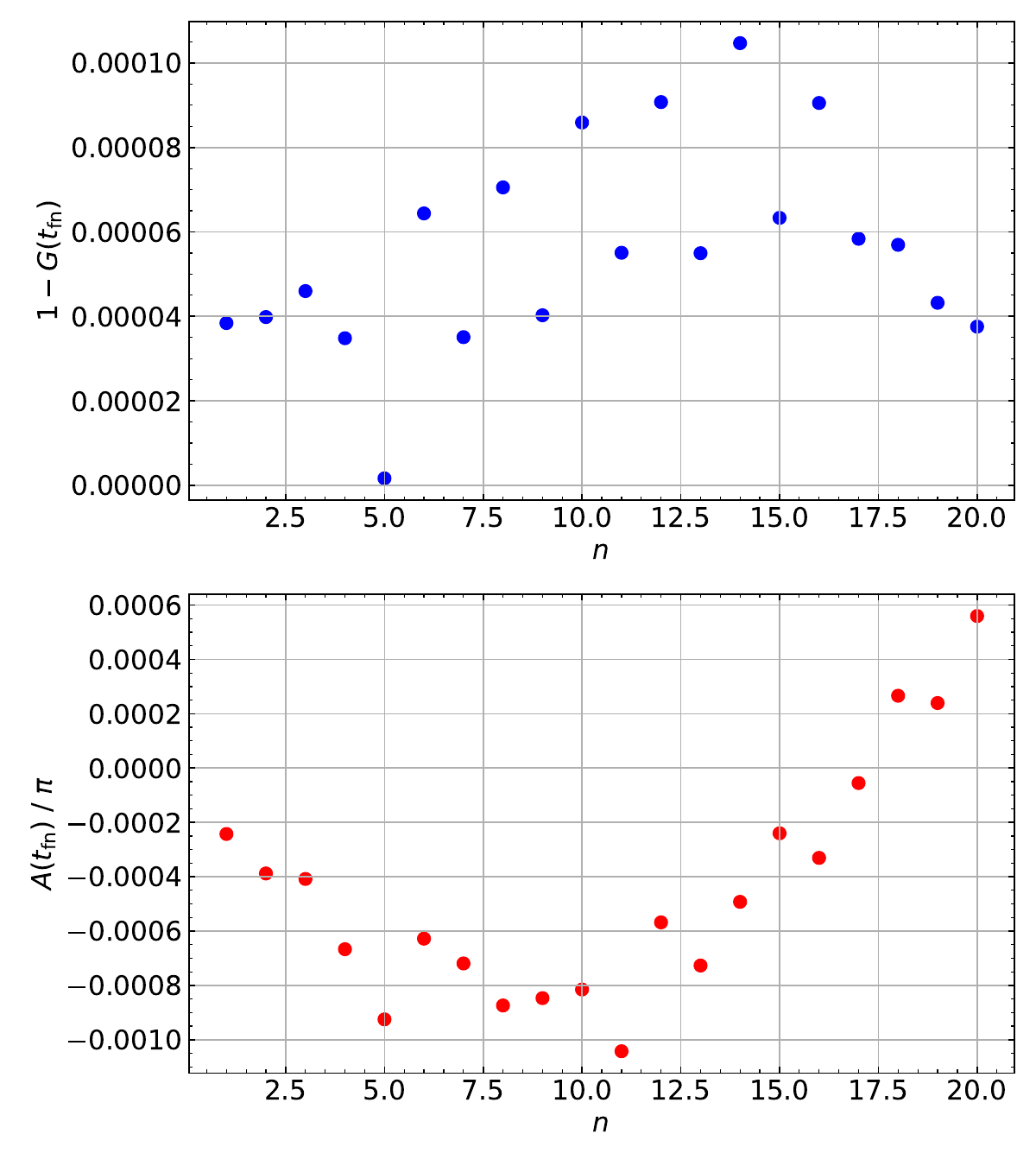}
\par\end{centering}
\caption{\label{fig:1-1}  The functions $G\left(t_{\mathrm{fn}}\right)$
and $A\left(t_{\mathrm{fn}}\right)$ involving three different
polarizations $i\in\left\{ x,y,z\right\} $ for the following parameters: $\omega=100\Er$, $\omega_\alpha=\Er$, $c=\kr$, $\tau=5\Er^{-1}$, $\tau^{\prime\prime}=5\Er^{-1}$, $\omega\Delta T=0.01$, $\theta_0=0$, $k_\beta=\kr$, $L=100\kr^{-1}$.}
\end{figure}

The functions $G\left(t_{\mathrm{fn}}\right)$ and $A\left(t_{\mathrm{fn}}\right)$ 
are presented in Fig.~\ref{fig:1-1}. One can see that $G\left(t_{\mathrm{fn}}\right)\approx1$
and $A\left(t_{\mathrm{fn}}\right)\approx0$.  This shows that  for $\theta = 0$ the overall dynamics from the initial to the final times is well described in terms of the effective dynamics governed by the zero order effective Hamiltonian.
 Indeed the first-order effective Hamiltonian $\tilde{H}_{\mathrm{eff}\left(1\right)}\left(t\right)$
presented in Eq.~(\ref{eq:H_eff1}) goes to zero for $\theta_{0}=\pi n_0$, where $n_0$ is an integer number.
Additionally, due to adiabatic ramping described by the ramping functions
$f\left(t\right)$ and $g\left(t\right)$, the effects of micro-motion
 disappear at the initial and and final times in the plots displayed in Fig.~\ref{fig:1-1} and in the subsequent Fig.~\ref{fig:1-1-1}. Therefore, for $\theta_{0}=\pi n_0$, the approximate dynamics
given by the zero-order effective Hamiltonian $\tilde{H}_{\mathrm{eff}\left(0\right)}\left(t\right)$,
is accurate up to second order in the inverse frequency.

\subsubsection{First-order correction effect}

\begin{figure}[]
\begin{centering}
\includegraphics[width=1\columnwidth]
{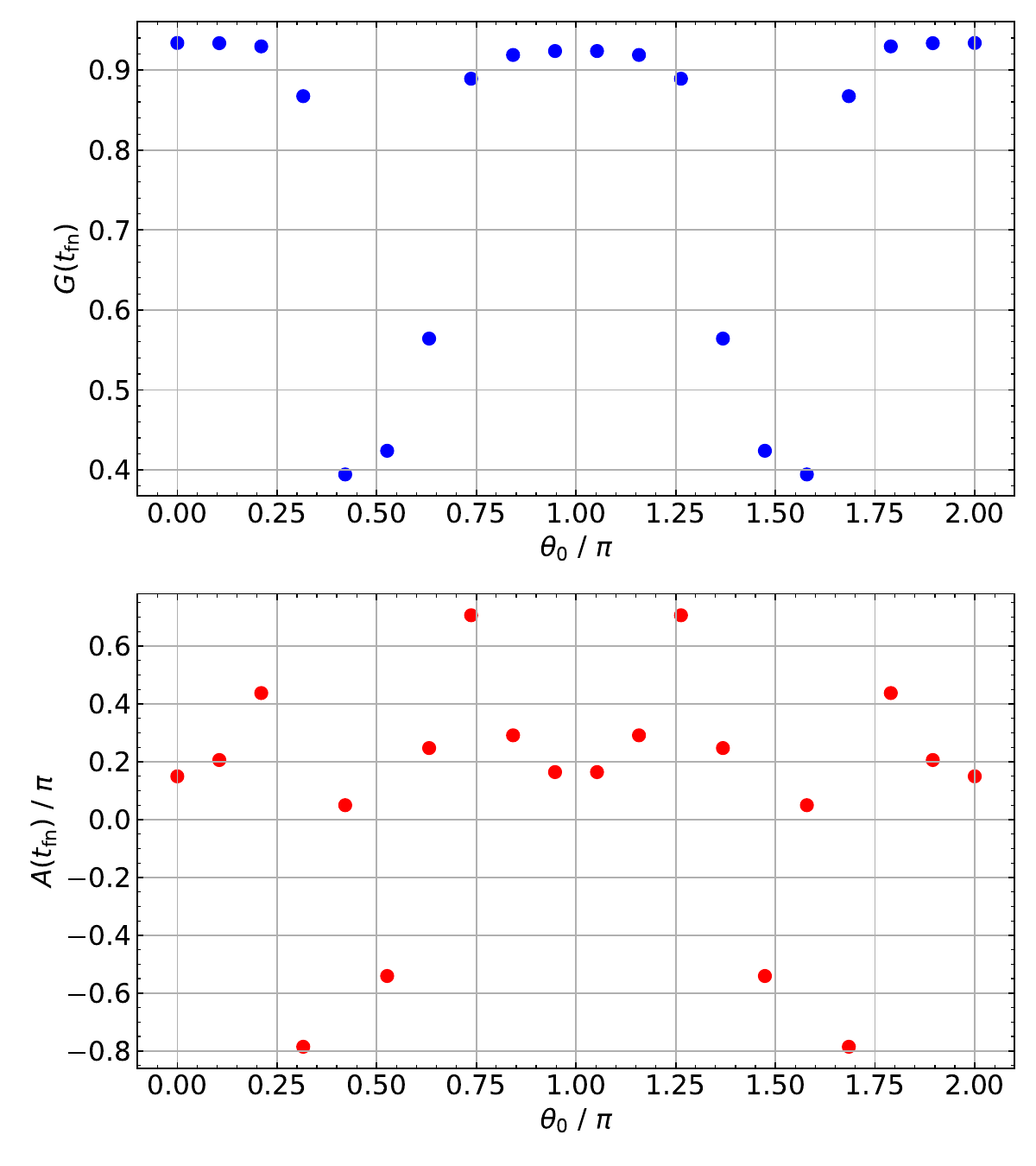}
\par\end{centering}
\caption{\label{fig:1-1-1} Comparison of dependence of functions $G\left(t_{\mathrm{fn}}\right)$
and $A\left(t_{\mathrm{fn}}\right)$ on $\theta_{0}$ for the following parameters: $\omega=10\Er$, $\omega_\alpha=\Er$, $c=0.1\Er$, $\tau=100\Er^{-1}$, $\tau^{\prime\prime}=150\Er^{-1}$, $\omega\Delta T=0.01$, $n=1$, $k_\beta=\kr$, $L=100\kr^{-1}$. }
\end{figure}

If $\theta_{0}\neq\pi n_0$, the first order effective Hamiltonian is no longer zero, and the approximate
dynamics is accurate only up to first order in the inverse frequency.  In Fig.~\ref{fig:1-1-1} we have demonstrate the difference in the approximation accuracy for various $\theta_{0}$ by calculating the dependence of
$G\left(t_{\mathrm{fn}}\right)$ on $\theta_{0}$. 
One can see clearly that the approximation holds best for $\theta_{0}\approx\pi n_0$ for which $\tilde{H}_{\mathrm{eff}(1)}=0$.
Here we have deliberately chosen the driving frequency $\omega$ to be considerably smaller than the one used in other plots, so that one can more clearly see the relative importance of first order correction term $\tilde{H}_{\mathrm{eff}(1)}$.

\section{Concluding remarks}

We have demonstrated how to by-pass the micro-motion emerging in the
magnetically induced SOC by switching on and off in a proper way the oscillating
magnetic fields at the initial and final times. We have studied the exact dynamics of the system
from the initial to the final times governed by the time periodic
Hamiltonian and compared it to the dynamics described by the slowly changing effective Floquet Hamiltonian. The two dynamics agree well under the assumption of the high frequency driving. The agreement is shown to be the best when the phase of the periodic driving takes a specific value for which the effect of the spin-orbit coupling is maximum. In that case the first-order effective Floquet vanishes and the zero-order Floquet Hamiltonian is correct up to the second order expansion in the inverse powers of the driving frequency. The overall dynamics is thus 
well described by the slowly
changing zero-order effective Floquet Hamiltonian containing the SOC term. In this way,
the magnetically induced SOC can be  induced in a controllable way
without involving the micro-motion. 
 This opens the path for practical applications of magnetically generated SOC, e.g. generation of nontrivial topological or spin-squeezed states for ultracold atoms in optical lattices, when the optically generated SOC is complicated to apply.

\begin{acknowledgments}
This work was supported by the European Social Fund (Project No. 09.3.3-LMT-K-712-23-0035) under a grant agreement with the Research Council of Lithuania (M.M.S.).
\end{acknowledgments}

\appendix

\section{Analysis of $\gamma\left(t\right)=\gamma\left(\omega t,t\right)$\label{sec:Analysis-of-gamma}}

\subsection{Function $\gamma\left(\omega t,t\right)$ \label{subsec:Function-gamma}}

Let us now find out how to separate a fast periodic time dependence
of $\gamma\left(t\right)=\gamma\left(\omega t,t\right)$ from its
additional slow temporal dependence. To that end, we expand $\gamma\left(t\right)$
as a series of $f^{(n)}\left(t\right)/\omega^{n}$ terms, where $f^{(n)}\left(t\right)$
denotes an $n$-th order temporal derivative of the slowly varying
envelope function $f\left(t\right)$. Substituting Eq.~(\ref{eq:beta})
into (\ref{eq:U_z-tilde}) and integrating by parts, one finds 
\begin{align}
\gamma\left(t\right)&=\omega\intop_{t_{\mathrm{in}}}^{t}f\left(s\right)\sin\left(\omega s-\theta_{0}\right)ds \nonumber \\
&=-\left.f\left(s\right)\cos\left(\omega s-\theta_{0}\right)\right|_{t_{\mathrm{in}}}^{t}+\left.\frac{f^{\prime}\left(s\right)}{\omega}\sin\left(\omega s-\theta_{0}\right)\right|_{t_{\mathrm{in}}}^{t}\nonumber \\
&-\intop_{t_{\mathrm{in}}}^{t}\frac{f^{\prime\prime}\left(s\right)}{\omega}\sin\left(\omega s-\theta_{0}\right)ds.\label{eq:gamma-integral}
\end{align}
This provides an expansion in a series of terms proportional to $f^{\left(n\right)}/\omega^{n}$:
\begin{align}
\gamma\left(t\right)&=\gamma\left(\omega t,t\right)\nonumber \\
&=-f\left(t\right)\cos\left(\omega t-\theta_{0}\right)+\frac{f^{\prime}\left(t\right)}{\omega}\sin\left(\omega t-\theta_{0}\right)\nonumber \\
&+\frac{f^{\prime\prime}\left(t\right)}{\omega^{2}}\cos\left(\omega t-\theta_{0}\right)+\ldots\,,\label{eq:gamma-rez-1}
\end{align}
where we used the fact that $f\left(t_{\mathrm{in}}\right)=f^{\prime}\left(t_{\mathrm{in}}\right)=f^{\prime\prime}\left(t_{\mathrm{in}}\right)=0$.

\subsection{Estimation of error\label{subsec:Estimation-of-error}}

To estimate an error made in writing Eq.~(\ref{eq:S_x-tilde_simplified})
for $\tilde{S}_{x}\left(z,t\right)$, let us expand the function $\gamma\left(\omega t,t\right)$
given by Eq.~(\ref{eq:gamma-rez-2}) in the powers of $t-t_{n}$
around a spike centered at $t_{n}=nT$, with an integer $n$. Since
amplitude $f\left(t\right)$ reaches its stationary value when $g\left(t\right)$
is still zero, one finds up to the quadratic term by putting $f\left(t\right)=1$:
\begin{equation}
\gamma\left(\omega t,t\right)\approx-\omega\left(t-t_{n}\right)\sin\theta_{0}+\omega^{2}\left(t-t_{n}\right)^{2}\cos\theta_{0}\,.\label{eq:gamma-Taylor}
\end{equation}
Therefore the maximum displacement $\left|t-t_{\mathrm{in}}^{\prime}\right|=\Delta T/2$
at which $\alpha\left(\omega t\right)$ is still non-zero yields the
following maximum value of $\left|\gamma\left(\omega t\right)\right|$:
\begin{equation}
\gamma_{\mathrm{max}}\approx\left|\left(\omega\Delta T/2\right)\sin\theta_{0}\right|+\left|\left(\omega\Delta T\right)^2/2\cos\theta_{0}\right|\,,\label{eq:gamma-Taylor-1}
\end{equation}
Since $\omega\Delta T=2\pi\Delta T/T\ll1$, then $\gamma_{\mathrm{max}}\ll1$. 

Equation (\ref{eq:S_x-tilde_simplified}) is valid if 
\begin{equation}
Lk_{\beta}\gamma_{\mathrm{max}}\ll1\,.\label{eq:Condition}
\end{equation}
where $L=z_{\mathrm{max}}$ is a characteristic size of the atomic
cloud. The conditions (\ref{eq:Condition}) holds best if the phase
difference is zero: $\theta_{0}=0$, i.e. when the spikes of the Zeeman
field $\alpha\left(\omega t\right)$ are situated at zero points of
the profile $\beta\left(\omega t\right)$. In that case $\gamma_{\mathrm{max}}=\left(\omega\Delta T\right)^{2}/2$
is quadratic in $\omega\Delta T$, and the condition (\ref{eq:Condition})
reduces to: 
\begin{equation}
Lk_{\beta}\left(\omega\Delta T\right)^{2}/2\ll1\,.\label{eq:Condition-1}
\end{equation}
Equations (\ref{eq:gamma-Taylor-1})-(\ref{eq:Condition}) or (\ref{eq:Condition-1})
provide restrictions on the size of the atomic cloud $L$. Since $Lk_{\beta}\gg1$,
the width of the spikes should be sufficiently small compared to the
driving period $T=2\pi/\omega$.

\section{First-order effective Hamiltonian \label{Appendix_B}
}

 Here will provide explicit calculations of the first-order effective Hamiltonian in the transformed frame.  The general formula for the first-order effective Hamiltonian is presented
by Eq.~\eqref{eq:H_eff1}:
\begin{equation}
\tilde{H}_{\textrm{eff}\left(1\right)}\left(t\right)=\frac{1}{\hbar\omega}\sum_{l=1}^{\infty}\frac{1}{l}\left[\tilde{H}^{\left(l\right)}\left(t\right),\tilde{H}^{\left(-l\right)}\left(t\right)\right]\,.
\label{H_eff1-Append}
\end{equation}
where $\tilde{H}^{\left(l\right)}\left(t\right)$ are the Fourier
components  of the transformed Hamiltonian  $\tilde{H}\left(\omega t,t\right)$ with respect to the first argument $\omega t$. 
 The latter $\tilde{H}\left(\omega t,t\right)$ is given by Eq.~\eqref{eq:H-tilde-1}:
\begin{align}
\tilde{H}\left(\omega t,t\right) & =H_{\mathrm{SI}}-\frac{p_{z}k_{\beta}}{m}S_{z}\gamma\left(\omega t,t\right)\nonumber \\
 & +\omega_{\alpha}g\left(t\right)\alpha\left(\omega t\right)\tilde{S}_{x}\left(z,t\right)+\frac{k_{\beta}^{2}}{2m}\gamma^{2}\left(\omega t,t\right)S_{z}^{2}\,.\label{eq:H-tilde-Appendix}
\end{align}

 Using the approximate expression \eqref{eq:gamma-rez-2-approx} for $\gamma\left(\omega t,t\right)$,
one has: $\gamma\left(\omega t,t\right)\approx-f\left(t\right)\cos\left(\omega t-\theta_{0}\right)$.
Thus  the non-zero Fourier modes of $\gamma\left(\omega t,t\right)$ with $m=\pm1$ read:
\begin{equation}
\gamma^{(\pm1)}\left(t\right)=-\frac{f\left(t\right)}{2}e^{\mp i\theta_{0}}\,.
\end{equation}
Since  the amplitude $\alpha\left(\omega t\right)$ is composed of sharp peaks  at $t=nT$,  the Fourier components $\alpha^{(\pm l)}$ weakly depend on $l$ and can be written: $\alpha^{(\pm l)}=1/2\pi$ for any $l\ge 0$.

  Next let us analyse the specific Fourier components $H^{(\pm l)}$  contributing to the effective Hamiltonian \eqref{H_eff1-Append}.

\subsection{ Contribution by $l=1$ Fourier modes}

Fourier components $\tilde{H}^{(l)}$ with  $l=\pm 1$ are:
\begin{equation}
\tilde{H}^{(\pm1)}\left(t\right)=\frac{k_{\beta}f\left(t\right)}{2m}e^{\mp i\theta_{0}}p_{z}S_{z}+\frac{\omega_{\alpha}g\left(t\right)}{2\pi}\tilde{S}_{x}\,.
\label{H^pm1-Appendix}
\end{equation}
The corresponding commutator  featured in the effective Hamiltonian \eqref{H_eff1-Append} reads:

\begin{equation}
\left[\tilde{H}^{\left(1\right)},\tilde{H}^{\left(-1\right)}\right]=-\frac{i\omega_{\alpha}k_{\beta}f\left(t\right)g\left(t\right)}{2\pi m}\sin\left(\theta_{0}\right)\left[p_{z}S_{z},\tilde{S}_{x}\right]\,.
\label{H^1,H^-1_Commutator-Appendix}
\end{equation}
The commutator may be rewritten as:

\begin{equation}
\left[p_{z}S_{z},\tilde{S}_{x}\right]=p_{z}\left[S_{z},\tilde{S}_{x}\right]+\left[p_{z},\tilde{S}_{x}\right]S_{z}\,.
\end{equation}
where 
\begin{equation}
\tilde{S}_{x}\left(z,t\right)=\cos\left(zk_{\beta}\cos\theta_{0}\right)S_{x}+\sin\left(zk_{\beta}\cos\theta_{0}\right)S_{y}\,.\label{eq:S_x-tilde_simplified-1}
\end{equation}
Using $\left[S_{z},\tilde{S}_{x}\right]=i\hbar\tilde{S}_{y}$ and $\left[p_{z},\tilde{S}_{x}\right]=-i\hbar k_{\beta}\cos\theta_{0}\tilde{S}_{y}$,
one obtains:

\begin{equation}
\left[p_{z}S_{z},\tilde{S}_{x}\right]=i\hbar p_{z}\tilde{S}_{y}-i\hbar k_{\beta}\cos\theta_{0}\tilde{S}_{y}S_{z}\,,
\end{equation}
where 
\begin{equation}
\tilde{S}_{y}=\cos\left(zk_{\beta}\cos\theta_{0}\right)S_{y}-\sin\left(zk_{\beta}\cos\theta_{0}\right)S_{x}\,.
\end{equation}
 In what follows we will consider the case of the spin-1/2. In that case one has $\tilde{S}_{y}S_{z}=i\hbar\tilde{S}_{x}/2$,
 so one can make further simplifications using $i\hbar k_{\beta}\cos\theta_{0}\tilde{S}_{x}=\left[p_{z},\tilde{S}_{y}\right]$.
 Consequently the commutator featured in Eq.~\eqref{H^1,H^-1_Commutator-Appendix} reduces to
\begin{equation}
\left[p_{z}S_{z},\tilde{S}_{x}\right]=i\frac{\hbar}{2}\left( p_{z}\tilde{S}_{y} + \tilde{S}_{y} p_{z} \right) \,.
\label{Commutator-Appendix}
\end{equation}
Substituting Eq.~\eqref{Commutator-Appendix} to Eqs.~\eqref{H^1,H^-1_Commutator-Appendix} and 
(\ref{H_eff1-Append}), one arrives at the first order effective Hamiltonian given by Eq.~\eqref{eq:H_eff1} in the main text.

\subsection{ Contribution by $l=2$ Fourier modes}

 Noting that:
\begin{equation}
\gamma^{2}\left(\omega t,t\right)\approx\frac{f^{2}\left(t\right)}{2}\left[1-\frac{1}{2}e^{-i2\theta_{0}}e^{i2\omega t}-\frac{1}{2}e^{i2\theta_{0}}e^{-i2\omega t}\right]\,,
\end{equation}
 the Fourier modes $\tilde{H}^{(l)}$ with  $l=\pm2 $ read:
\begin{equation}
\tilde{H}^{(\pm2)}\left(t\right)=\frac{\omega_{\alpha}g\left(t\right)}{2\pi}\tilde{S}_{x}-\frac{k_{\beta}^{2}f^{2}\left(t\right)}{32m}e^{\mp i2\theta_{0}}S_{z}^{2}\,.
\label{H^pm2-Appendix}
\end{equation}
 For spin-1/2 one has $S_{z}^{2}=1/4$, so the last term
of Eq.~\eqref{H^pm2-Appendix} is proportional to the identity
operator,  and the commutator $\left[\tilde{H}^{\left(2\right)}\left(t\right),\tilde{H}^{\left(-2\right)}\left(t\right)\right]$
reduces to zero. For  arbitrary spin,
the commutator$\left[\tilde{H}^{\left(2\right)}\left(t\right),\tilde{H}^{\left(-2\right)}\left(t\right)\right]$
is no longer zero and the first-order effective Hamiltonian would be more
complicated. 

\subsection{ Contribution by Fourier modes with $l>2$}

 The Fourier $H^{(\pm l)}\left(t\right)$ with $l>2$  
are all the same:
\begin{equation}
H^{(\pm l)}\left(t\right)=\frac{\omega_{\alpha}g\left(t\right)}{2\pi}\tilde{S}_{x}\,\quad \mathrm{for}\quad l>2\,.
\end{equation}
 so the commutators $\left[H^{\left(l\right)}\left(t\right),H^{\left(-l\right)}\left(t\right)\right]$
vanish for $l>2$.

\subsection{Final result}

In this way the first-order effective Hamiltonian reads using Eqs.~\eqref{H_eff1-Append}, \eqref{H^1,H^-1_Commutator-Appendix} and \eqref{Commutator-Appendix}:
\begin{equation}
H_{{\rm eff}\left(1\right)}\left(t\right)=\frac{\omega_{\alpha}\hbar k_{\beta}f\left(t\right)g\left(t\right)}{4\pi m\hbar\omega}\sin\left(\theta_{0}\right)\left( p_{z}\tilde{S}_{y} + \tilde{S}_{y} p_{z} \right)\,.
\label{H_eff1-final-Appendix}
\end{equation}

\bibliography{SOC-Magnet-grad-2017}

\end{document}